\documentstyle[12pt]{article}

\begin{document}
\pagestyle{myheadings}

\newcommand{\pstar}{\mbox{$\psi^{\ast}$}}

\newcommand{\be}{\begin{equation}}
\newcommand{\ee}{\end{equation}}
\newcommand{\bea}{\begin{eqnarray}}
\newcommand{\eea}{\end{eqnarray}}

\title{Post-Gaussian Variational Method for the Nonlinear Schrodinger Equation:
Soliton Behavior and  Blowup}

\author {Fred Cooper \\
{\small \sl Theoretical Division, Los Alamos National Laboratory,}\\
{\small \sl Los Alamos, NM 87545}\\
\\
Harvey Shepard \\
{\small \sl Physics Department, University of New Hampshire,}
{\small \sl Durham, NH 03824}\\
\\
Carlo Lucheroni and Pasquale Sodano\\
{\small \sl Dipartimento di Fisica and Sezione INFN Universita di Perugia
,}
{\small \sl 06100 Perugia,Italy       }}\\

 \maketitle

\begin{abstract} We use Dirac's time-dependent variational principle to discuss
several features of the general nonlinear Schrodinger equation

$
i {\frac{\partial \psi}{\partial t}} + \nabla^{2} \psi +
|\pstar\psi|^{\kappa}  \psi = 0
$
in $d$ spatial dimensions for arbitrary nonlinearity parameter $\kappa$. We
employ a family of trial variational wave functions, more general than
Gaussians, which can be treated analytically and which preserve the
canonical  structure (and hence the conservation laws) of the exact system.
As examples, we  derive an approximation to the one-dimensional soliton
solution and demonstrate  the ``universality'' of the critical exponent for
blowup in the supercritical  case, $\kappa d > 2$. For the critical
case $\kappa d = 2$, we find that one gets an excellent estimate for the
critical mass necessary for blowup when we minimize the blowup mass with
respect to the non-gaussian variational parameter.
\end{abstract}

\section{Introduction and General Formalism} In a recent work \cite{CLS:p}
we showed how Dirac's time-dependent variational principle could be used to
study self-focusing in the general class of nonlinear Schrodinger equations:

\be
i {\frac{\partial \psi}{\partial t}} + \nabla^{2} \psi +
|\pstar\psi|^{\kappa}  \psi = 0
\ee
in $d$ spatial dimensions for arbitrary nonlinearity parameter $\kappa$.
Gaussian trial variational wave functions were used to obtain qualitative
information about blowup at finite times as a function of $d$ and $\kappa$. For
the critical case, $\kappa d=2$, we derived an analytic expression for the
critical mass necessary for blowup that is in good agreement with numerical
results in one and two dimensions. For the supercritical case, $\kappa d > 2$,
we determined the critical exponents as a function of $\kappa$ and $d$.

In the present work we extend the formalism by considering a more general class
of variational wave functions and illustrate their usefulness by ({\sl 1})
showing how the one-dimensional soliton solution can be approximated, and ({\sl
2}) demonstrating ``universality''; viz., that the critical exponent for
finite-time blowup in the supercritical case is identical for all of our trial
functions. We also find that in the critical case the mass necessary for
blowup depends on the power $2n$ in the exponent of
our trial wave function.By minimizing the mass with respect to $n$ we
get a dramatic improvement in our estimate of the blowup mass when
compared with our previous Gaussian trial wave function estimate.  In
a separate work we apply the extended class of variational  functions to
several
quantum-mechanical problems \cite{CS:p}.

 The  time-dependent variational
principle, posited by Dirac \cite{Dirac:p}, has been  found very useful for
various quantum mechanics and field theory problems
\cite{JK:p},\cite{KK:p},\cite{CPS:p},\cite{CM:p},\cite{PS:p}.  Since the
nonlinear Schrodinger equation (NLSE) has a  Hamiltonian and a conserved
probability, with probability density given by   $P(x,t) = \pstar(x,t)
\psi(x,t)$, the same variational principle can be used to  derive the NLSE as
well as for obtaining time- dependent approximate variational  solutions.
The
starting point for the derivation of the NLSE is Dirac's  variational principle
\cite{Dirac:p}. The action for the NLSE is

\be
\Gamma = \int dt L   \label{eq:gamma-definition}
\ee
where L is given by

\be
L = {\frac{i}{2}} \int d^d x (\pstar \psi_{t} -\pstar_{t}\psi) - H
\ee
For the NLSE in $d$ dimensions with nonlinearity parameter $\kappa$, we have
\cite{Das:p}:

\be
H = \int d^{d} x [\nabla\pstar \nabla\psi -
{\frac{(\pstar\psi)^{\kappa+1}}  {\kappa+1}}]
\ee
The NLSE with arbitrary nonlinearity follows from Dirac's variational principle
that $\Gamma$ be stationary against variations in $\psi(x,t)$ and
$\pstar(x,t)$;
that is, ${\frac{\delta\Gamma}{\delta\psi}} =
{\frac{\delta\Gamma}{\delta\pstar}} = 0$ implies

\be i \psi_{t} + \nabla^{2}\psi +  (\pstar\psi)^{\kappa} \psi = 0 \ee

\be i \pstar_{t} - \nabla^{2}\pstar -  (\pstar\psi)^{\kappa} \pstar = 0 \ee

$\Gamma$ is the effective action of the system described by $\psi(x,t)$ and the
variational principle is a version of Hamilton's least-action principle. In the
variational principle $\psi$ is an arbitrary square-integrable function subject
to the conservation law

\be {\frac{d}{dt}} (\int d^{d} x [\pstar\psi]) = 0. \ee

To obtain an approximate solution to the NLSE we consider a restricted class of
$\psi(x,t) = \psi_{v}(x,t)$ where $\psi_{v}(x,t)$ is constrained to be of a
form
that tries to capture the known behavior of the full $\psi$ for the problem at
hand. In the present work we consider a general class of variational wave
equations that includes the Gaussian as a special case:

\be
\psi_{v}(x,t) = A(n,d,G_{2n}) \exp \{-|x-q(t)|^{2n} [\frac{d}{4nG_{2n}(t)} -
i\Lambda (t)]+ ip(t)[x-q(t)] \} \ee

Thus, in the most general case we have four real time-dependent variational
parameters: $p(t$), $q(t)$, $G(t)$, and $\Lambda(t)$. As we shall see, in many
cases of interest this number must be reduced in order to preserve the
canonical
structure (and thereby maintain the conservation laws). There are a number of
obvious advantages to this general class of variational functions: ({\sl 1})
applications to a broad range of initial conditions are possible; ({\sl 2}) by
having a whole family of trial functions in addition to the Gaussian for
which  analytic calculations can be performed, one can study the
sensitivity of the  variational estimates to the choice of function; ({\sl
3}) unlike the $n=1$  Gaussian case, the other wave functions have non-zero
higher order correlation  functions.

\section{Equations of Motion}

We shall consider the general $d$-dimensional ($d > 1$) spherically symmetric
case and then indicate
the (minor) changes needed for the $d=1$ case. For $d>1$, we must restrict our
class of variational functions to those for which $p(t) = q(t) = 0$. We write
(with $r = |{\bf x}|$)

\be  \psi_{v} =\left (\frac{2nM}{\Omega_{d} \Gamma
(\frac{d}{2n})}\right)^{1/2} (\frac{2nG_{2n}(t)}{d})^{-d/4n} \exp
[-r^{2n}(\frac{d}{4nG_{2n}(t)} - i\Lambda (t))]    \label{eq:psi_v}
\ee

where

\be \Omega_{d} \equiv \frac{2\pi ^{d/2}}{\Gamma(d/2)} \ee

\be d^{d}x = \Omega_{d} r^{d-1} dr \ee

then $\psi_{v}$ is normalized such that

\be
M = \int_{0}^{\infty} \psi^{\ast}_{v} \psi_{v} d^{d}x
\ee

is a constant. Also,

\be
G_{2n} = \langle r^{2n} \rangle
\ee

where

\be
\langle f \rangle \equiv \int f \frac{\psi^{\ast}_{v} \psi_{v}}{M} d^{d}x
\ee

Calculating the pieces of the action for the class of trial wave functions
(\ref{eq:psi_v}), we find

\bea
 \frac{i}{2} \int (\pstar\psi_{t} - \pstar_{t}\psi)
d^{d}x   = -\dot{\Lambda} \int \pstar\psi r^{2n} d^{d}x   = - \dot{\Lambda} M
\langle r^{2n} \rangle = -\dot{\Lambda} G_{2n} M.
\label{eq:piece-of-the-action}
\eea

Using

\be
\langle r^{\alpha} \rangle = \int_{0}^{\infty} r^{\alpha} (\frac
{\pstar\psi}{M}) \Omega_{d} r^{d-1} dr =\left
(\frac{2nG_{2n}}{d}\right)^{\alpha/2n}
{\frac{\Gamma[\frac{\alpha+d}{2n}]}{\Gamma[\frac{d}{2n}]}},
\ee

we have

\bea
 H_{\rm free} = \int \nabla\pstar \cdot \nabla\psi d^{d}x =
4n^{2}[(\frac{d}{4nG_{2n}})^{2} + \Lambda^{2}] M \langle r^{4n-2}
\rangle\nonumber \nonumber \\= 4n^{2}[(\frac{d}{4nG_{2n}})^{2} + \Lambda^{2}]
(\frac{2nG_{2n}}{d})^{\frac{4n-
2}{2n}} {\frac{\Gamma[2+\frac{d-2}{2n}]}{\Gamma[\frac{d}{2n}]}}
\label{eq:H_free}
\eea

and

\bea
 H_{\rm int} = -(\frac{1}{\kappa+1}) \int
(\pstar\psi)^{\kappa+1}d^{d}x   = -(\frac{1}{\kappa+1}) M \int
(\pstar\psi)^{\kappa} (\frac{\pstar\psi}{M})d^{d}x \nonumber \\  = -
(\frac{1}{\kappa+1}) M \langle  (\pstar\psi)^{\kappa} \rangle
 = -M [\frac{2nM}{\Omega_{d}\Gamma[\frac{d}{2n}]}]^{\kappa}
(\frac{2nG_{2n}}{d})^{-
\kappa d/2n} (1+\kappa)^{-1-d/2n}.    \label{eq:H_int}
\eea

Defining

\be
\Gamma_{R} \equiv {\frac{\Gamma[2+\frac{d-2}{2n}]}{\Gamma[\frac{d}{2n}]}},
\ee

and abbreviating $G \equiv G_{2n}(t)$, we use equations
(\ref{eq:gamma-definition}), (\ref{eq:piece-of-the-action}), (\ref{eq:H_free}),
and (\ref{eq:H_int}) to
form the action:

\bea  \Gamma = M \int dt \{-G \dot{\Lambda} - 4n^{2}[\Lambda^{2} +
(\frac{d}{4nG})^{2}] (\frac{2nG}{d})^{2-\frac{1}{n}}\Gamma_{R}  \nonumber \\
    +   (1+\kappa)^{-1-d/2n}  (\frac{2nG}{d})^{-\kappa d/2n}
[\frac{2nM}{\Omega_{d}\Gamma[\frac{d}{2n}]}]^{\kappa} \}\label{eq:action}
\eea
This form of the effective action, $\Gamma$, guarantees that
$\frac{dH}{dt} = 0 $ follows from the variational equations of
motion:

\be
 {\frac{\delta\Gamma}{\delta\Lambda}} = 0 = \dot{G}(t) -
8n^{2} (\frac{2nG}{d})^{2- {1/n}}\Gamma_{R}\Lambda(t)   \label{eq:eqn-motion-1}
\ee

or

\be
\Lambda(t) = \dot{G}(t) (\frac{2nG}{d})^{{1/n}-2} (8n^{2}\Gamma_{R})^{-1}
\label{eq:Lambda}
\ee
and
\bea
{\frac{\delta\Gamma}{\delta G}} = 0 = -\dot{\Lambda}(t) - 4n^{2}\Gamma_{R}
(\frac{2n}{d})^{2-{1/n}} \{(2-1/n) G^{1-{1/n}}
[\Lambda^{2}+(\frac{d}{4nG})^{2}] \nonumber \\
     - 2G^{-1-{1/n}}(\frac{d}{4n})^{2} \}
- \kappa (\frac{2n}{d})^{-{\frac{\kappa d}{2n}}-1}
(1+\kappa)^{-1-
{\frac{d}{2n}}} (\frac{2nM}{\Omega_{d}\Gamma[\frac{d}{2n}]})^{\kappa} G^{-1-
{\frac{\kappa d}{2n}}}.    \label{eq:eqn-motion-2}
\eea

These equations of motion can be combined and integrated to yield a statement
of
conservation of energy, which we can write directly from (\ref{eq:H_free}) and
(\ref{eq:H_int})

\bea
\widehat{E} \equiv \frac{(H_{\rm free} + H_{\rm int})}{M} =  4n^{2}
(\frac{2nG}{d})^{2-{1/n}}
[\Lambda^{2}+(\frac{d}{4nG})^{2}]\Gamma_{R} \nonumber \\
    - (1+\kappa)^{-1-{\frac{d}{2n}}}
(\frac{2nM}{\Omega_{d}\Gamma[\frac{d}{2n}]})^{\kappa} [\frac{2nG}{d}]^{-
\frac{\kappa d}{2n}}.    \label{eq:Energy}
\eea

Combining (\ref{eq:Lambda}) and (\ref{eq:Energy}), we have

\bea
\widehat{E} = n^{2} \Gamma_{R}  (\frac{2nG}{d})^{-{1/n}} +
\dot{G}^{2}(\frac{2nG}{d})^{{1/n}-2} [16n^{2}\Gamma_{R}]^{-1} \nonumber \\
     -  (1+\kappa)^{-1-{\frac{d}{2n}}}
(\frac{2nM}{\Omega_{d}\Gamma[\frac{d}{2n}]})^{\kappa} [\frac{2nG}{d}]^{-
\frac{\kappa d}{2n}}.    \label{eq:Energy2}
\eea

For the $d=1$ case there are some adjustments. Now we can include nonzero
variational parameters $p(t)$ and $q(t)$ and still preserve the canonical
structure.

The class of variational trial functions is

\be  \psi_{v} = (2nG(t))^{-\frac{1}{4n}}
(\frac{nM}{\Gamma[\frac{1}{2n}]})^{1/2} \exp[-|x- q(t)|^{2n}
[{\frac{1}{4nG(t)}}
- i\Lambda(t)] + ip(t)[x-q(t)]] \label{eq:L}
\ee

where as above

\be M = \int_{-\infty}^{\infty} \pstar_{v}\psi_{v} dx \ee

and

\be G = \langle |x-q|^{2n} \rangle. \ee

Evaluating the effective action, we find

\be  \Gamma = M \int dt [p(t)\dot{q}(t) -p^{2}(t)] + { \Gamma\{\rm
eq.}(\ref{eq:action})\}_{d=1}.   \label{eq:M}
\ee

The variational equations

\be {\frac{\delta\Gamma}{\delta\Lambda}} = 0  \; {\rm and}  \;
{\frac{\delta\Gamma}{\delta G}} = 0 \ee

yield (\ref{eq:eqn-motion-1}) and (\ref{eq:eqn-motion-2}), evaluated at $d=1$.
In
addition, we now have variational equations for $p(t)$ and $q(t)$:

\be
{\frac{\delta\Gamma}{\delta q}} = 0 = \dot{p} \Longrightarrow p(t)
= {\rm const.} \equiv p_{0}      \label{eq:N}
\ee

\be
{\frac{\delta\Gamma}{\delta p}} = 0 \Longrightarrow \dot{q}(t) =
2p(t) = 2p_{0}.     \label{eq:O}
\ee

The conserved energy is given by (\ref{eq:Energy}) or (\ref{eq:Energy2}), with
$d=1$
and $E$ replaced by $E-p_{0}^{2}$.


\section{Applications}

\subsection{Solitons}

We define soliton solutions as those for which the width
\be
G(t)  = constant
\ee
Hence   $ \dot{G}(t) = 0 $ and then eq. (\ref{eq:Lambda})
implies $\Lambda(t) = \dot{\Lambda} = 0 $ . Using these values, and solving
(\ref{eq:eqn-motion-2}) for $G$,

\bea
(2nG/d)=
(1+\kappa)^\frac{d+2n}{2-\kappa d}
\left[\frac{2nM}{\Omega_{d}\Gamma[\frac{d}{2n}]}\right]^
\frac{2n\kappa}{\kappa d -2}
\left[\frac{\kappa d}{2n^{2}\Gamma_{R}}\right]^\frac{2n}{\kappa d -2}
\label{eq:G1} . \eea

Substituting (\ref{eq:G1}) into (\ref{eq:Energy}), with $\Lambda = 0 $ yields

\bea
\widehat{E} = p_{0}^{2}
+\left[(\frac{2nM}{\Omega_{d}\Gamma[\frac{d}{2n}]})\right]^{2\kappa/(2-\kappa
d)}
(1+\kappa)^{2(1+d/2n)/(\kappa d - 2)} \times \nonumber \\
\left( n^{2} \Gamma_{R}
\left[\frac{d \kappa}{2n^{2}\Gamma_{R}}\right]^{\frac{2}{2-\kappa d}}
- \left[\frac{d \kappa}{2n^{2}\Gamma_{R}}\right]^{\frac{\kappa d}{2-\kappa
d}} \right)   \label{eq:E3},
\eea
where $ p_{0} = 0 $ for $ d \neq 1 $.

The single soliton solution to the usual one-dimensional
NLSE is well known.
 As given in \cite{Drazin:b}, one has:
\be
\phi = (M/2\sqrt{2}) \exp [i (v/2)(x-vt)] sech [(M/4)(x-vt)]
\ee
where $\phi$ has been normalized as in eq. (27), and $v=2p_{0}$.

To compare with this solution we
 consider the one-dimensional $
d = \kappa = 1 $ case.  Then $\widehat{E}$, eq.(\ref{eq:E3}), becomes
\be
\widehat{E} = p_{0}^2 - ( M^2/16) [2^{1/n} \Gamma (1/2n) \Gamma
(2-1/2n)]^{-1} \ee
The behavior of $\widehat{ E} $ as a function of $n$ is shown in Fig. 1. We
see that  all the trial wave functions have energy higher than that of the
exact single soliton which has
\be
 \widehat{E} = E/M = p_{0}^{2} - {\frac {1}{48}} M^2.
\ee
 We see that
$\widehat{E}$ is minimized for $n\approx 0.8 $ At this value (37)
coincides with (38).
 The variational
wave function, eq.(\ref{eq:L}) for $ \kappa = d=1$
is
\be
\psi_{v} = \left[ {\frac{n M}{\Gamma(1/2n)}}\right]^{1/2}(2n G(t))^{-1/4n}
 \exp\left[- (4 n G(t))^{-1}
 |x-vt |^{2n} + i {\frac {v}{2}} (x-vt)\right]      \label{eq:psisol}
\ee

where, from eq. (34),
$$
G= 2^{4n} n^{2 n-1} \left[ \Gamma(2-1/2n) \right]^{2n} M^{-2n}
$$

The best approximate wave function is that found by the energy minimization
procedure.
In Fig. 2 we plot $ | \psi_{v}|^2 $ and $|\phi|^2$ for $M=1$ showing how
well the non-Gaussian variational function with n=4/5 approximates the
soliton  solution.
In Fig. 3 we plot  the difference  $ | \psi_{v}|^2 - |\phi|^2$

In Fig. 4 we plot the same comparison as in fig. 2 for n=1(Gaussian trial
functions) and for  n=1/2.

\subsection{ Blowup and Universality }
First, we demonstrate that the rate of
finite-time blowup for  the supercritical case is independent of $n$ in the
class of variational trial  functions.
As in \cite{CLS:p} we discuss blowup (self-focusing) by
requiring that for small width $G(t)$, it is possible to
have $ \dot{G}(t) < 0 $ so that $ G(t) \rightarrow 0 $ in a
finite time.

Solving eq. (\ref{eq:Energy2} ) for $ \dot{G}(t) $, we find
that
$\dot{G}$ obeys an equation of the form:

\bea
 \dot{G}^2 = E A(n,d) (2nG/d)^{\frac {4 n -2}{2 n}} + M^{\kappa}
 C(\kappa,n,d)
(2nG/d)^{\frac {4 n -2 -d \kappa}{2 n}}  \nonumber \\
 - D(n,d) (2nG/d)^{\frac {4 n -4}{2
n}}. \label {eq:bup}
\eea

where
$$
A (n,d) = 16n^{2} \Gamma_{R}, \;  D(n,d) = 16n^{4} \Gamma_{R}^{2}
$$

$$
C(\kappa,n,d)=16n^{2} \Gamma_{R} (1+\kappa)^{-1-d/2n}
(\frac{2n}{\Omega_{d}\Gamma[\frac{d}{2n}]})^{\kappa}
$$
 From this equation we
see that if $ \kappa d  > 2$  (supercritical case) only the second term is
important and blow up always occurs at a finite time  $ t^{\star}$ ,
whereas when $ \kappa d = 2 $ (critical case) the
second and third  terms have the same power of G and lead to there being
a critical mass necessary for blowup.
 In the ``supercritical'' case
\be
\dot{G}(t) = - f(n,d,\kappa) G^{(4n-2-\kappa d)/4n},
\ee
which implies that
\be
G(t) \propto (t^{\ast} - t ) ^ {4n/(\kappa d + 2)}
\ee
and hence $G(t) $ reaches zero at some finite time $t^{\ast}
$.
 From eq. (\ref{eq:psi_v}) ,
$ \psi \propto G^{-d/4n} $
and thus
\be
\psi \propto  (t^{\ast} - t ) ^ {-d/(\kappa d + 2)}
\ee
with a critical exponent independent of $n$.

Next let us turn to the critical case:
 $\kappa  d = 2$ . From eq.(\ref{eq:bup}) we
see that for small G only the last two terms only  are important. So that

\be
 \dot{G}^2 = \left[ M^{\kappa}
 C(\kappa,n,d) - D(n,d) \right] (2nG/d)^{\frac {4 n -4}{2
n}}.
\ee

The condition for blowup is that $ \dot{G} < 0 $ for small G. This
requires that

\be
M > \left[{\frac{D(n,d)} {C (n,d)}} \right]^ {d/2}  \label{eq:mc}
\ee

Thus in our approximation the critical mass is clearly a function of the
shape parameter ``n''. Eq.( \ref{eq:mc}) has the property that the
critical mass is a minimum at a particular value of n.  In Fig. 5 we plot
the critical mass for d=1 as a function of the variational parameter n.
We find that at n=.727, M= 2.746 which is to be compared with the numerical
value 2.7 \cite{Rose:p} and the analytic value of 2.8569 for n=1, the
Gaussian case. Thus we see that by adding an extra variational parameter n
we dramatically improve our estimate for the critical mass.

In Fig. 6 we plot the critical mass for d=2 as a function of the variational
parameter n . Here we see that the best value of n is .693 for which
M= 11.84 which is to be compared with the numerical value 11.73
\cite{Schochet:p} and the previous gaussian result (n=1) of 12.56. Again we
see the dramatic  improvement of our estimate for the blowup mass.

In Fig. 7 we plot the critical mass for d=3 as a function of the variational
parameter n. Here we see the best n is around .667 giving a best value
of M= 64.56 instead of our previous gaussian estimate of M=69.5.

Thus we find going beyond a gaussian  trial wave function gives us
a better approximate soliton and an accurate estimate of the critical
mass for blowup. We also find that the critical exponent does not
depend on the new parameter n.

\section*{Acknowledgments}
We thank Eliot Shepard for help in preparing the manuscript. One of us
(F.C.) would like to thank both the University of New Hampshire and
the University of Perugia for  their hospitality. This work was supported
in part by grants from the DOE and the INFN.

\end{document}